\documentclass[10pt,superscriptaddress,twocolumn,amsmath,amssymb,aps,pra,showpacs]{revtex4-1}
\usepackage{mathrsfs}
\usepackage{graphicx}
\usepackage{dcolumn}
\usepackage{bm}
\usepackage[title,titletoc,header]{appendix}

\usepackage{amssymb}
\usepackage{amsmath}
\usepackage{mathrsfs}
\usepackage{amsmath}
\usepackage{subfigure}
\usepackage{float}
\usepackage[title,titletoc,header]{appendix}
\usepackage{graphicx}

\newcommand{\be}{\begin{equation}}
\newcommand{\ee}{\end{equation}}
\newcommand{\bea}{\begin{eqnarray}}
\newcommand{\eea}{\end{eqnarray}}

\begin{document}
\title{Infinite bound states and $1/n$  energy spectrum induced by a Coulomb-like potential of type III in a flat band system }
\author{Yi-Cai Zhang}
\affiliation{School of Physics and Materials Science, Guangzhou University, Guangzhou 510006, China}



\date{\today}
\begin{abstract}
In this work, we investigate the bound states in  a one-dimensional  spin-1  flat band system  with a Coulomb-like potential of type III, which has a unique non-vanishing matrix element in basis $|1\rangle$.
It is found that,  for such a kind of potential, there exists infinite bound states.
Near the threshold of continuous spectrum, the bound state energy is consistent with the ordinary hydrogen-like atom energy level formula with Rydberg correction.
In addition, the flat band has significant effects on the bound states.
For example, there are infinite bound states which are generated from the flat band.
Furthermore, when the potential  is weak, the bound state energy is proportional to the Coulomb-like potential strength $\alpha$.
When the bound state energies are very near the flat band, they are inversely proportional to the natural number $n$ (e.g., $E_n\propto 1/n, n=1,2,3,...$).
Further we find that the energy spectrum can be well described by quasi-classical approximation (WKB method). Finally, we give a critical potential strength $\alpha_c$ at which  the bound state energy reaches the threshold  of continuous spectrum. \textbf{After crossing the threshold, the  bound states in the continuum (BIC) may exist in such a flat band system.}


\end{abstract}

\maketitle
\section{Introduction}
A lot of  novel physical phenomena, for example, existences of localized flat band states \cite{Sutherland1986,Vidal1998,Mukherjee}, the ferro-magnetism transition \cite{Mielke1999,Zhang2010,Raoux2014}, localization \cite{Leykam2017}, super-Klein tunneling \cite{Shen2010,Urban2011,Fang2016,Ocampo2017}, quantum Hall effects \cite{Yang2012}, Zitterbewegung \cite{Ghosh}, preformed pairs \cite{Tovmasyan2018}, strange metal \cite{Volovik2019}, superconductivity/superfluidity \cite{Peotta2015,Hazra2019,Cao2018,Wuyurong2021,Kopnin2011,Julku2020,Iglovikov2014,Julku2016,Liang2017,Iskin2019,Wu2021}, ect., can  appear in a flat band system.
One of the most prominent features of flat band is that the existence of an infinitely large density of states. It is well known that the behaviors of density of states near the threshold of a continuous energy spectrum play crucial roles in the formations of bound states \cite{Economou}. In a spin-1 flat band system, due to  the peculiar density of states and its ensuing $1/z$ singularity of Green function, a short-ranged potential, e.g., square well potential,  can result in infinite bound states, even a hydrogen atom-like  energy spectrum, i.e., $E_n\propto1/n^2,n=1,2,3,...$ \cite{Zhangyicai2021}.

In addition, it is found that the existences of  bound states also depend on the types of potentials.
 For example, in the presence of  a long-ranged Coulomb-like potential of type I (with three same diagonal elements in usual basis),
an arbitrarily weak Coulomb-like potential can destroy completely the flat band in two-dimensional spin-1 model \cite{Gorbar2019}.
Further more, Pottelberge found that the flat band gradually  evolves into a continuous band  with the increasing of Coulomb-like potential strength \cite{Pottelberge2020}.
In two-dimensional flat band systems, a strong Coulomb-like potential can result in a wave function collapse near the the origin \cite{Gorbar2019,Han2019}. For one-dimensional case, an arbitrarily weak Coulomb-like potential also  causes the wave function collapse  \cite{Zhangyicai20212}.

For a potential of type II, which has a unique non-vanishing matrix element in  basis $|2\rangle$ \cite{Zhangyicai2021}, a short-ranged potential, e.g., square well potential, can cause an infinite number of  bound states, even a hydrogen atom-like energy spectrum.
In addition, such a kind of Coulomb-like potential can result in a $1/n$ energy spectrum \cite{Zhangyicai20212}.

In this work, we investigate the bound states in a one-dimensional spin-1 Dirac-type Hamiltonian with a  Coulomb-like potential of type III, which have only one non-vanishing matrix element in basis $|1\rangle$.
It is found that,  depending on the sign of potential strength over bound state energy, i.e., $\alpha/E$ , there exist two different effective potentials.
When $\alpha/E<0$, the effective potential has a lowest point in coordinate space. The bound states can exist in the whole space. In addition,  there are infinite bound states which are generated from a continuous energy spectrum. Near the threshold of continuous energy spectrum, the  bound state energy is reduced to the ordinary hydrogen atom energy spectrum.
When $\alpha/E>0$, the effective potential has no lowest point.
   Similarly as that in Coulomb-like potential of type II \cite{Zhangyicai20212}, there are also infinite bound states which are generated from the flat band. Near the the flat band, the energy is inversely proportional to the natural number, i.e., $E\propto1/n$.
Differently from the ordinary one-dimensional bound state energy which is a parabolic function of potential strength,  the bound state energy is linearly dependent on the potential strength as the strength goes to zero. For a given quantum number $n$, the bound state energy grows up with the increasing of potential strength $\alpha$.  We give a critical potential strength $\alpha_c$ at which  the bound state energy reaches the threshold  of continuous spectrum. After crossing the threshold, the bound states may still exist in the continuous spectrum, which indicates that the  bound states in a continuum (BIC) may exist in such the flat band system.


The work is organized as follows. In Sec.\textbf{II}, the three energy bands, free particle wave functions are given.  Next, we solve the bound state problem for a Coulomb-like potential of type III in Sec.\textbf{III}.
 At the end, a summary is given in Sec.\textbf{IV}.

\section{The model Hamiltonian  with a flat band}
In this work, we consider a spin-1 Dirac-type Hamiltonian \cite{Zhangyicai2021} in one dimension, i.e.,
 \begin{align}
&H=H_0+V_p(x)\notag\\
&H_0=-iv_F\hbar S_x\partial_x+m S_z,
    	\label{hamiltonian}
\end{align}
where $V_p(x)$ is potential energy, $H_0$ is the free-particle Hamiltonian, $v_F>0$ is Fermi velocity, and $m>0$ is energy gap parameter. $S_x$ and $S_z$ are spin operators for spin-1 particles \cite{Zhang2013}, i.e.,
\begin{align}
&S_x=\left[\begin{array}{ccc}
0 &\frac{1}{\sqrt{2}}  & 0\\
\frac{1}{\sqrt{2}}& 0 &\frac{1}{\sqrt{2}}\\
0 &\frac{1}{\sqrt{2}} & 0
  \end{array}\right];&S_z=\left[\begin{array}{ccc}
1 &0  & 0\\
0&0& 0\\
0 &0 & -1
  \end{array}\right],
\end{align}
in usual basis $|i\rangle$ with $i=1,2,3$. In the whole manuscript, we use the units of $v_F=\hbar=1$. The above Hamiltonian can be realized in photonic systems \cite{Huang2011,Chan2012}.
 When $V_p(x)=0$, the free particle Hamiltonian $H_0$ has three eigenstates  and the
 eigenenergies, i.e.,
\begin{align}\label{3}
&\langle x |-,k\rangle=\psi_{-,k}(x)=\frac{1}{2\sqrt{k^2+m^2}}\left(\begin{array}{ccc}
\sqrt{k^2+m^2}-m\\
 -\sqrt{2}k\\
\sqrt{k^2+m^2}+m
  \end{array}\right)e^{ikx},\notag\\
  &E_{-,k}=-\sqrt{k^2+m^2};\notag\\
  &\langle x |0,k\rangle=\psi_{0,k}(x)=\frac{1}{\sqrt{2(k^2+m^2)}}\left(\begin{array}{ccc}
-k\\
 \sqrt{2}m\\
k
  \end{array}\right)e^{ikx},\notag\\
  &E_{0,k}=0;\notag\\
  &\langle x |+,k\rangle=\psi_{+,k}(x)=\frac{1}{2\sqrt{k^2+m^2}}\left(\begin{array}{ccc}
\sqrt{k^2+m^2}+m\\
 \sqrt{2}k\\
\sqrt{k^2+m^2}-m
  \end{array}\right)e^{ikx},\notag\\
  &E_{+,k}=\sqrt{k^2+m^2}.
\end{align}
It is found that a flat band with zero energy ($E_{0,k}=0$) appears in between upper and lower bands (see Fig.1). The possible bound states may exist in the gaps among the three bands, i.e., $0<E<m$ and $-m<E<0$ (the regions \textbf{A} and \textbf{B} in the Fig.1). Some localized flat band states can be obtained by superpositions of the above wave functions $\psi_{0,k}(x)$ \cite{Zhangyicai20212}. The localized flat band wave functions show a logarithmic singularity near their center positions.
\begin{figure}
\begin{center}
\includegraphics[width=1.0\columnwidth]{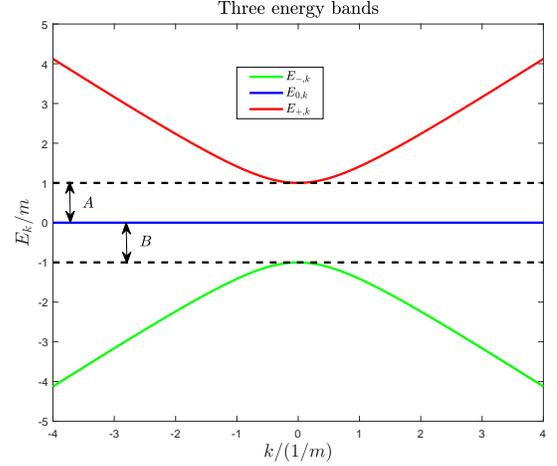}
\end{center}
\caption{ The energy spectrum of free particle Hamiltonian.
  The possible bound states only exist in the regions \textbf{A} and \textbf{B}. }
\label{schematic}
\end{figure}

\section{bound states in a Coulomb-like potential of type III}
In the following manuscript, we assume the potential energy $V_p$ has following form in usual basis $|i=1,2,3\rangle$, namely,
\begin{align}\label{V}
&V_p(x)=V_{11}(x)\bigotimes|1\rangle\langle1|
=\left[\begin{array}{ccc}
V_{11}(x) &0  & 0\\
0&0& 0\\
0 &0 & 0
  \end{array}\right].
\end{align}
In the whole manuscript, we would refer such a kind of potential as potential of type III.\textbf{ Such a spin-dependent
potential is a bit similar to the magnetic impurity potential in Kondo model, which may be realized in flat band materials of solid physics. }
The similar bound state problems with potential of type I and II have been investigated by Zhang and Zhu \cite{Zhangyicai2021,Zhangyicai20212}.

Now the Schr\"{o}dinger  equation with three component wave functions can be written as
\begin{align}\label{5}
&-i\partial_x\psi_2(x)/\sqrt{2}=[E-m-V_{11}]\psi_1(x),\notag\\
&-i\partial_x[\psi_1(x)+\psi_3(x)]/\sqrt{2}=E\psi_2(x),\notag\\
&-i\partial_x\psi_2(x)/\sqrt{2}=[E+m]\psi_3(x).
\end{align}

Eliminating  $\psi_2$ and $\psi_3$,  we get an equation for $\psi_1$
\begin{align}\label{H0}
-\partial_{x}^{2}[\frac{E-V/2}{E+m}\psi_{1}(x)]=E(E-m-V)\psi_1(x).
\end{align}

Further we introduce a new wave function $\psi(x)\equiv\frac{E-V/2}{E+m}\psi_{1}(x)$, then the equation for $\psi(x)$ is
\begin{align}\label{H0}
\partial_{x}^{2}\psi(x)+\frac{E(E-m-V)(E+m)}{E-V/2}\psi(x)=0.
\end{align}
Let's write it into a form of an effective Schr\"{o}dinger equation (a second-order differential equation), i.e.,
\begin{align}\label{eff}
-\partial_{x}^{2}\psi(x)+\tilde{V}\psi(x)=\tilde{E}\psi(x).
\end{align}
where effective total energy $\tilde{E}$ and effective potential $\tilde{V}$ are
\begin{align}\label{H0}
&\tilde{E}=E^2-m^2<0, \ for\ ordinary \ bound  \ states,\notag\\
&\tilde{V}=\frac{V_{11}}{2}\frac{(m+E)^2}{E-V_{11}/2}.
\end{align}
In the determining the effective total energy, we assume that $V_{11}(x)\rightarrow0$ as $x\rightarrow\pm\infty$.

Next we assume $V_{11}$ is  a Coulomb-like potential, i.e.,
\begin{align}\label{H0}
V_{11}(x)=\frac{\alpha}{|x|},
\end{align}
where $\alpha$ describes the potential strength.
The effective potential $\tilde{V}$ is
\begin{align}\label{11}
\tilde{V}=\frac{V_{11}}{2}\frac{(m+E)^2}{E-V_{11}/2}=\frac{A}{|x|-x_0}.
\end{align}
In the above equation, we introduce parameter $A\equiv\frac{\alpha (m+E)^2}{2E}$ and $x_0\equiv\frac{\alpha}{2E}$. It is shown that the effective potential $\tilde{V}$ is a shifted Coulomb-like potential with an effective potential strength $A$ \cite{Downing2014}, which depends on energy $E$.
The Eq.(\ref{eff}) becomes
\begin{align}\label{Coul}
\partial_{x}^{2}\psi(x)+[\tilde{E}-\frac{A}{|x|-\frac{\alpha}{2E}}]\psi(x)=0.
\end{align}
In the following, we would solve the effective Schr\"{o}dinger equation Eq.(\ref{Coul}) to get the bound state energies.

For $x>0$, Eq.(\ref{Coul}) can be solved with some confluent hypergeometric functions. Its general solution is
\begin{align}
&\psi(x)=(x-x_0)e^{-\sqrt{-\tilde{E}}(x-x_0)}\{c_{1}\times {}_1F_1[a,b,2\sqrt{-\tilde{E}}(x-x_0)]\notag\\
&+c_{2}\times U[a,b,2\sqrt{-\tilde{E}}(x-x_0)]\}
\end{align}
where ${}_1F_1[a,b,z]=\sum_{k=0}^{\infty}\frac{(a)_kz^k}{k!(b)_k}$ is confluent hypergeometric function,  $(a)_k=a\times (a+1)\times(a+2)\times...\times(a+k-1)$, and $c_1(c_2)$ are two arbitrary constants. $a=1+\frac{A}{2\sqrt{-\tilde{E}}}$, $b=2$.
When $z\rightarrow+\infty$, ${}_1F_1(a,b,z)$ has an asymptotic expansion  \cite{Abramowitz}
\begin{align}\label{23}
&{}_1F_1[a,b,z]= \frac{\Gamma(b)}{\Gamma(a)}e^{z}z^{a-b}[1+O(1/|z|)],
\end{align}
where $\Gamma(a)$ is the Euler Gamma function.
 $U[a,b,z]$ is a second linearly independent solution to the confluent hypergeometric equation (Tricomi function \cite{Wang1989}),
 whose asymptotic expansion  is \cite{Abramowitz}
\begin{align}\label{24}
&U[a,b,z]= (\frac{1}{z})^a[1+O(1/|z|)]
\end{align}
as $z\rightarrow \infty$.
When $z\rightarrow 0$, the two confluent hypergeometric functions behave as
\begin{align}\label{25}
&{}_1F_1[a,b,z]\simeq1,\notag\\
&U[a,b,z]= \frac{\Gamma(b-1)}{\Gamma(a)}z^{1-b}+O(|logz|), \quad (b=2).
\end{align}
When $a$ is real, $b$ is an integer,  and $z<0$, the imaginary part of $U[a,b,z]$ is proportional to ${}_1F_1[a,b,z]$, i.e.,
\begin{align}\label{16}
&Im\{U[a,b,z]\}= \frac{\pi(-1)^b}{(b-1)!\Gamma(a-b+1)}{}_1F_1[a,b,z],
\end{align}
where $Im\{U\}$ is the imaginary part of  $U$. When $a$ is real, $b$ is an integer,  and $z>0$, $Im\{U[a,b,z]\}\equiv0$.
In addition, when the parameter $a\rightarrow \infty$ and $z>0$, the confluent hypergeometric function ${}_1F_1[a,b,-z/a]$ and  $U[a,b,-z/a]$  would be transformed into Bessel functions \cite{Abramowitz}, i.e.,
\begin{align}\label{261}
&lim_{a\rightarrow\infty}{}_1F_1[a,b,-z/a]=\Gamma(b)z^{1/2-b/2}J_{b-1}(2\sqrt{z}),\notag\\
&lim_{a\rightarrow\infty}U[a,b,-z/a]=\frac{-\pi ie^{i\pi b}z^{1/2-b/2}H^{(1)}_{b-1}(2\sqrt{z})}{\Gamma(1+a-b)},\notag\\
\end{align}
where $J_{\nu}(x)$ and $H^{(1)}_{\nu}(x)$ are the $\nu-$th order Bessel function and Hankel function of first kind, respectively.
The above properties of confluent hypergeometric functions would be very useful in the following discussions.

\begin{figure}
\begin{center}
\includegraphics[width=1.0\columnwidth]{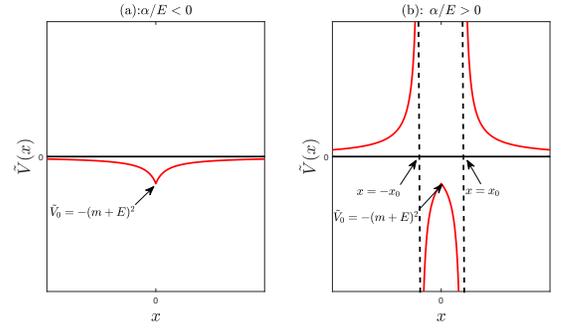}
\end{center}
\caption{ Two kinds of effective potentials (the red solid lines). (a): The effective potential for $\alpha/E<0$. The effective potential has a lowest point $\tilde{V}_0=-(m+E)^2$ at origin $x=0$ (b): The effective potential for $\alpha/E>0$. The value of effective potential at $x=0$, i.e., $\tilde{V}_0=-(m+E)^2$, is not a lowest point. For $m>E>0$, the effective total energy $\tilde{E}>\tilde{V}_0$. While for $-m<E<0$, the effective total energy $\tilde{E}<\tilde{V}_0$.  }
\label{schematic}
\end{figure}

Depending on the sign of $\alpha/E$, there exists two kinds of effective potentials $\tilde{V}$ (see Fig.2).
 For usual bound states, the energy should satisfy $0<E<m$ or $-m<E<0$.
 When $\alpha/E<0$,   the effective potential $\tilde{V}$ has a lowest point at $x=0$, i.e., $\tilde{V}_{0}=-(m+E)^2$ (see Fig.2).
 In addition, the effective total energy $\tilde{E}=E^2-m^2$  should be larger than the lowest point of potential, i.e, $\tilde{E}-\tilde{V}_{0}=2E(m+E)>0$, then $E>0$.
So the bound states energy should be larger than zero for the case of  $\alpha/E<0$ (see Fig.3).

When $\alpha/E>0$, the effective potential $\tilde{V}$ has no lowest point (see Fig.2). The effective potential $\tilde{V}$ is negative in the interval $(-x_0,x_0)$, and positive in intervals $(-\infty,-x_0)$ and $(x_0,\infty)$. There are two infinitely high potential barriers near two ends  $x=\pm x_0$ of the interval $(-x_0,x_0)$.
 It is found that the bound state energy can be larger than zero for  $\alpha>0$ or  smaller than zero for  $\alpha<0$ (see Figs.4 and 5).

\subsection{$\alpha/E<0$}
When $\alpha/E<0$ [$x_0=\alpha/(2E)<0$], the bound states can  exist in the whole space ($-\infty,\infty$). At the two ends ($x=\pm\infty$), the zero boundary conditions should be satisfied, i.e.,
\begin{align}
&\psi(\pm\infty)=0.
\end{align}
Considering  Eqs.(\ref{23}) and (\ref{24}), ${}_1F_1[a,b,2\sqrt{-\tilde{E}}(x-x_0)]$ should be discarded. Then, the wave function is
\begin{align}
&\psi(x)=(x-x_0)e^{-\sqrt{-\tilde{E}}(x-x_0)}U[a,b,2\sqrt{-\tilde{E}}(x-x_0)].
\end{align}

In addition, due to the presence of parity symmetry ($x\rightarrow-x$), the wave functions can be classified by two distinct parities, i.e., odd and even parities. For odd parity states, the wave functions at the origin $x=0$ should vanish.
While for even parity states, the derivatives of the wave functions at the origin $x=0$ are zero.
Based on these boundary conditions,  we get the bound state energy equations
\begin{align}
&\psi(x=0)=0,\ for\ odd \ parity \ states,\notag\\
&\psi'(x=0)=0,\ for\ even \ parity \ states.
\end{align}
To be specific, for odd parity states, the bound state energy equation is
\begin{align}\label{29}
&U[1+\frac{\alpha(E+m)^2}{4E\sqrt{m^2-E^2}},2,-\frac{\alpha\sqrt{m^2-E^2}}{E}]=0.
\end{align}
For even parity states, the bound state energy equation is
\begin{align}\label{30}
&U[\frac{\alpha(E+m)^2}{4E\sqrt{m^2-E^2}},0,-\frac{\alpha\sqrt{m^2-E^2}}{E}]\notag\\
&-2U[\frac{\alpha(E+m)^2}{4E\sqrt{m^2-E^2}},1,-\frac{\alpha\sqrt{m^2-E^2}}{E}]=0.
\end{align}
The results are reported in Fig.(3).
\begin{figure}
\begin{center}
\includegraphics[width=1.0\columnwidth]{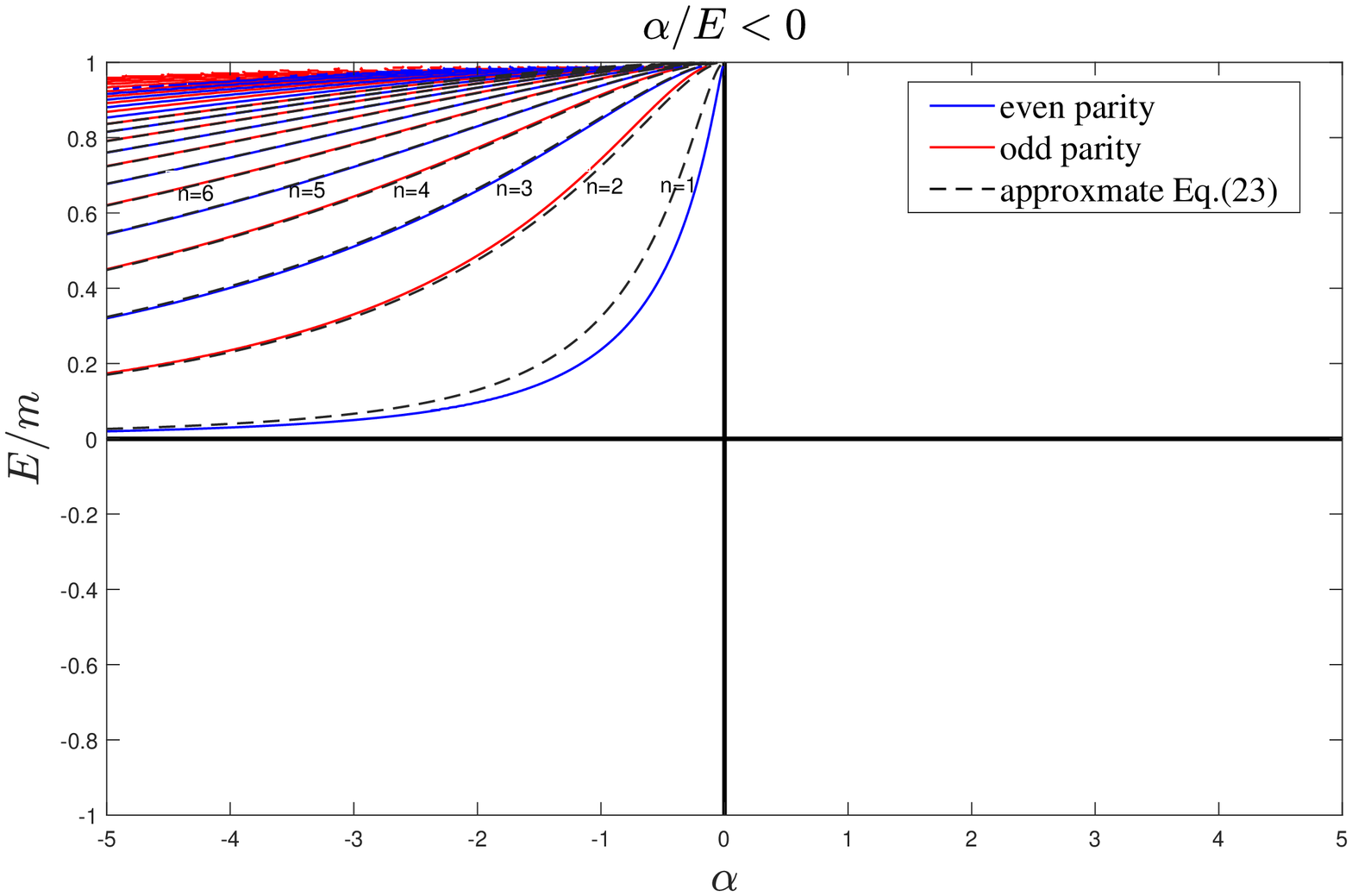}
\end{center}
\caption{ The bound state energy of Coulomb-like potential in the case of $\alpha/E<0$. The solid lines are the the exact results of Eqs.(\ref{29}) and (\ref{30}). The black dashed lines are given by the quasi-classical approximation formula Eq.(\ref{31}). }
\label{schematic}
\end{figure}

The bound state energy can be also given by the quasi-classical approximation \cite{Landau}, i.e., Wenzel-Kramers-Brillouin (WKB) method.  With the quasi-classical approximation, the energy is
\begin{align}\label{31}
&\frac{-\alpha\sqrt{E(E+m)}}{2E}\{-\sqrt{2}+(2\sqrt{\frac{E}{m-E}}+\sqrt{\frac{m-E}{E}})\notag\\
&\times arctan[\sqrt{\frac{2E}{m-E}}]\}=(n+\Delta)\pi,
\end{align}
where quantum number $n=1,2,3,...$, and
\begin{align}\label{321}
&\Delta=-\frac{1}{4},\ for\ odd \ parity \ states,\notag\\
&\Delta=-\frac{3}{4},\ for\ even \ parity \ states.
\end{align}
Fig.3 shows that the bound state energy can be well described by the quasi-classical approximation formula Eq.(\ref{31}).

When the energy  approaches the threshold $m$ of upper  energy band , i.e., $|E-m|/m\ll1$, it can be approximated by
\begin{align}\label{27}
E_B\equiv E_n\equiv E\simeq m[1-\frac{\alpha^2}{2(n+\Delta)^2}]
\end{align}
where $\Delta$ is  the Rydberg correction of hydrogen-like atom energy level \cite{Landau}, which  takes same values as Eq.(\ref{321}) and $n=1,2,3,...$.
When $n\gg1$, the bound state energy  (relative to the threshold $m$)
 \begin{align}\label{27}
E_n-m\simeq -\frac{m \alpha^2}{2n^2},
\end{align}
which is consistent with the hydrogen atom energy levels \textbf{of s-states}.
It is because when $n\gg1$, the particles in these  bound states are far away from the origin, and the effective potential can be viewed as ordinary Coulomb potential, i.e., $\tilde{V}\simeq A/|x|$ [see Eq.(\ref{11})].
When $\alpha\rightarrow -\infty$, for a given $n$, we find the energy
$E\propto 1/\alpha^2\rightarrow0$.

\subsection{$\alpha/E>0$}
When $\alpha/E>0$[$x_0=\alpha/(2E)>0$], due to the existence of the infinitely high  potential barriers near $x=\pm x_0$, there exist two different cases.
\subsubsection{The bound states only exist in the interval $(-x_0,x_0)$ }
For such a case, outside the interval $(-x_0,x_0)$, the wave function vanishes.
At the two ends of the interval, the zero boundary conditions should be satisfied, i.e.,
\begin{align}\label{281}
&\psi(\pm x_0)=0.
\end{align}
Taking  Eq.(\ref{25}) into account, $U[a,b,2\sqrt{-\tilde{E}}(x-x_0)]$ should be discarded. So the wave function is
\begin{align}\label{227}
&\psi(x)\notag\\
&=(x-x_0)e^{-\sqrt{-\tilde{E}}(x-x_0)}{}_1F_1[a,b,2\sqrt{-\tilde{E}}(x-x_0)],  \ |x|\leq x_0,\notag\\
&\psi(x)= 0, \ \ \ \ |x|>x_0.
\end{align}

Similarly, the wave functions can be classified by parities.
For odd parity states, the bound state energy equation is
\begin{align}\label{361}
{}_1F_1[1+\frac{\alpha(E+m)^2}{4E\sqrt{m^2-E^2}},2,-\frac{\alpha\sqrt{m^2-E^2}}{E}]=0.
\end{align}
For even parity states, the bound state energy equation is
\begin{align}\label{37}
&[-4E(2E+\alpha\sqrt{m^2-E^2})]\notag\\
&\times{}_1F_1[1+\frac{\alpha(E+m)^2}{4E\sqrt{m^2-E^2}},2,-\frac{\alpha\sqrt{m^2-E^2}}{E}]\notag\\
&+\alpha[4 E\sqrt{m^2-E^2}+\alpha(m+E)^2]\notag\\
&\times{}_1F_1[2+\frac{\alpha(E+m)^2}{4E\sqrt{m^2-E^2}},3,-\frac{\alpha\sqrt{m^2-E^2}}{E}]=0.
\end{align}
The results are reported in Fig.(4).

For $E<0$,  with quasi-classical approximation method, the energy is given by
\begin{align}\label{38}
\frac{\alpha(m+E)^{3/2}\pi}{4E\sqrt{m-E}}=n\pi,
\end{align}
where $n=1,2,3,...$.
When $\alpha\rightarrow-\infty$, the energy $E\rightarrow -m$.
It is found that the odd and even parity states have approximately same energies for a given $n$ (see Fig.4).
This is because when $E<0$, the effective total energy $\tilde{E}=E^2-m^2$ is smaller than the potential energy at the origin, i.e., $\tilde{E}-\tilde{V}_0=E^2-m^2+(E+m)^2=2E(m+E)<0$.
The origin $x=0$ belongs to the classically forbidden region, then the values of wave functions near the origin would be very small. Consequently, the two boundary conditions for odd and even parity bound states, i.e., $\psi(0)=0$ and $\psi'(0)=0$, are basically equivalent, and then the bound state energies are doubly degenerate approximately for a given $n$.

\begin{figure}
\begin{center}
\includegraphics[width=1.0\columnwidth]{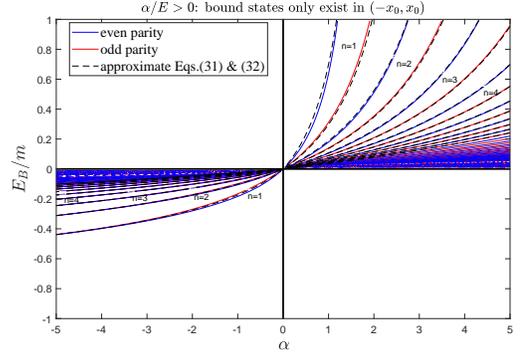}
\end{center}
\caption{ The bound state energy of Coulomb-like potential in the case of $\alpha/E>0$. The solid lines are the the exact results of Eqs.(\ref{361}), and (\ref{37}) . The black dashed lines are given by the quasi-classical approximation formulas Eqs.(\ref{38}), and (\ref{32}). }
\label{schematic}
\end{figure}

When $E>0$, with quasi-classical approximation method,  the eigen-energy  is given by
\begin{align}\label{32}
&\frac{\alpha(m+E)}{2E}[\frac{\sqrt{2E}}{\sqrt{m+E}}+\frac{\sqrt{m+E}}{\sqrt{m-E}}arcsin(\frac{\sqrt{m-E}}{\sqrt{m+E}})]\notag\\
&=(n+\Delta)\pi,
\end{align}
where $n=1,2,3,....$, and
\begin{align}\label{33}
&\Delta=+\frac{1}{4},\ for\ odd \ parity \ states,\notag\\
&\Delta=-\frac{1}{4},\ for\ even \ parity \ states.
\end{align}
When $E/m\ll1$, the energy can be approximated by
\begin{align}
E_B=E_n=E\simeq\frac{m\alpha}{4(n+\Delta)},
\end{align}
where $n=1,2,3,...$, $\Delta$ takes same values as Eq.(\ref{33}). When $n\gg1$, the bound state energy is
\begin{align}
&E_n\simeq\frac{m\alpha}{4n}\propto1/n.
\end{align}
It indicates that near the flat band, the bound state energies are proportional to potential strength $\alpha$, and  they are inversely proportional to the natural number $n$ for large quantum number, which are similar to the case of Coulomb-like potential of type II \cite{Zhangyicai20212}.

For a given quantum number $n$, the bound state energy grows up with the increasing of potential strength $\alpha$ (see Fig.4). When $\alpha$ reaches a critical value $\alpha_c$, the bound state energy $E$ would reach the threshold of upper continuous spectrum, i.e., $E=m$.
When energy  approaches the threshold $m$, the parameter $a=1+\frac{A}{2\sqrt{-\tilde{E}}}=1+\frac{\alpha(E+m)^2}{4E\sqrt{m^2-E^2}}\rightarrow \infty$.
Using Eq.(\ref{261}), the Eq.(\ref{361}) and Eq.(\ref{37}) can be represented  by
\begin{align}\label{36}
&J_{1}(2\alpha_c)=0, \ for\ odd \ parity \ states,\notag\\
&\alpha J_{2}(2\alpha_c)-J_{1}(2\alpha_c)=0,\ for\ even \ parity \ states.
\end{align}
It should be emphasized that Eq.(\ref{36}) is exact result for $E=m$.

Further using the asymptotic formula of Bessel functions, i.e, $J_{\nu}(x)\sim \sqrt{\frac{2}{\pi x}}cos(x-\nu\pi/2-\pi/4)$ as $x\rightarrow\infty$ , the critical potential strength $\alpha_c$ can be approximated by
\begin{align}\label{27}
&\alpha_c\simeq\frac{(1/4+n)\pi}{2},\ for\ odd \ parity \ states,\notag\\
&\alpha_c\simeq\frac{(-1/4+n)\pi}{2},\ for\ even \ parity \ states,
\end{align}
where natural number $n=1,2,3,...$.
After crossing these critical values, the bound states may still exist and they would form the bound states in a continuum (BIC). We would give detailed discussions on the existence of bound states in the continuum (BIC) elsewhere \cite{BIC}.

\subsubsection{The bound states exist in the whole space $(-\infty,\infty)$ }
For such a case,  outside the interval $(-x_0,x_0)$, the wave function does not vanish.
In addition, at the two ends of the interval $(-x_0,x_0)$, the wave function takes some finite values, i.e.,
\begin{align}\label{271}
&\psi(\pm x_0)\neq0.
\end{align}
Using Eqs.(\ref{23}-\ref{16}), it is found that the wave function can be represented with
\begin{align}\label{228}
&\psi(x)=(x-x_0)e^{-\sqrt{-\tilde{E}}(x-x_0)}Re\{U[a,b,2\sqrt{-\tilde{E}}(x-x_0)]\}.
\end{align}
where $Re\{U\}$ is the real part of $U$.

Similarly, the wave functions can be classified by parities.
For odd parity states, the bound state energy equation is
\begin{align}\label{391}
Re\{U[1+\frac{\alpha(E+m)^2}{4E\sqrt{m^2-E^2}},2,-\frac{\alpha\sqrt{m^2-E^2}}{E}]\}=0.
\end{align}
For even parity states, the bound state energy equation is
\begin{align}\label{401}
&Re\{U[\frac{\alpha(E+m)^2}{4E\sqrt{m^2-E^2}},0,-\frac{\alpha\sqrt{m^2-E^2}}{E}]\notag\\
&-2U[\frac{\alpha(E+m)^2}{4E\sqrt{m^2-E^2}},1,-\frac{\alpha\sqrt{m^2-E^2}}{E}]\}=0.
\end{align}
The results are reported in Fig.(5).

Similarly, for $E<0$,  with quasi-classical approximation method, the energy is given by
\begin{align}\label{41}
\frac{\alpha(m+E)^{3/2}\pi}{4E\sqrt{m-E}}=(n-1/2)\pi,
\end{align}
where $n=1,2,3,...$.

\begin{figure}
\begin{center}
\includegraphics[width=1.0\columnwidth]{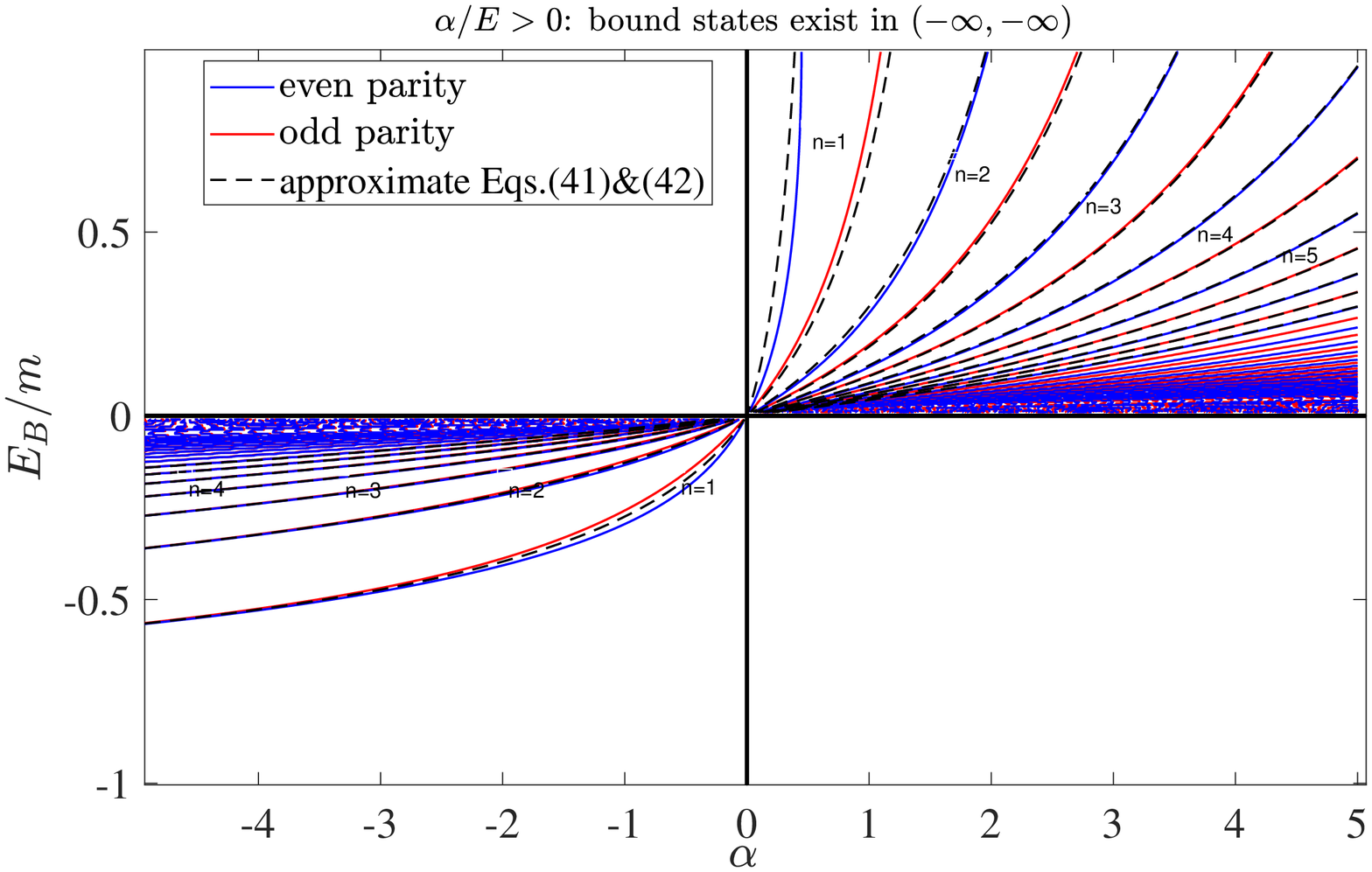}
\end{center}
\caption{ The bound state energy of Coulomb-like potential in the case of $\alpha/E>0$. The solid lines are the the exact results of Eqs.(\ref{391}), and (\ref{401}) . The black dashed lines are given by the quasi-classical approximation formulas Eqs.(\ref{41}), and (\ref{42}). }
\label{schematic}
\end{figure}

\begin{figure}
\begin{center}
\includegraphics[width=1.0\columnwidth]{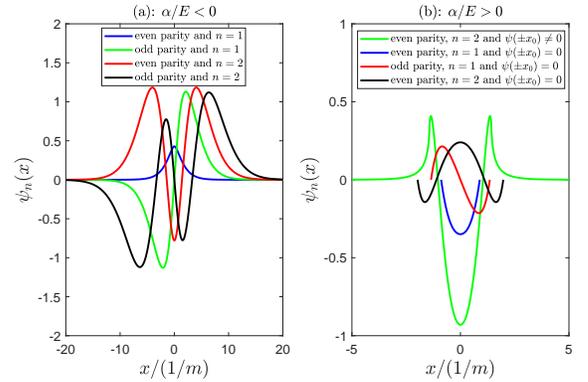}
\end{center}
\caption{ \textbf{The wave functions (un-normalized). Here we take the same bound state energy $E=0.5m$ for all the wave functions. (a): The wave functions for  $\alpha/E<0$. (b): The wave functions for  $\alpha/E>0$.
The green solid line of panel (b) is a wave function described by Eq.(40). Other three lines correspond to the formula Eqs.(29). } }
\label{schematic}
\end{figure}

When $E>0$, with quasi-classical approximation method,  the eigen-energy  is given by
\begin{align}\label{42}
&\frac{\alpha(m+E)}{2E}[\frac{\sqrt{2E}}{\sqrt{m+E}}+\frac{\sqrt{m+E}}{\sqrt{m-E}}arcsin(\frac{\sqrt{m-E}}{\sqrt{m+E}})]\notag\\
&=(n+\Delta)\pi,
\end{align}
where $n=1,2,3,....$, and
\begin{align}\label{441}
&\Delta=-\frac{1}{4},\ for\ odd \ parity \ states,\notag\\
&\Delta=-\frac{3}{4},\ for\ even \ parity \ states.
\end{align}
Comparing Fig.5 with Fig.4, we see they have similar energy spectra. However,  there is a $\pi/2$ phase difference in quasi-classical approximation wave functions between the two cases [see also Eqs.(\ref{38}) and  (\ref{41}) or Eqs.(\ref{33}) and  (\ref{441})]. This is because the boundary conditions of the wave functions at $x=\pm x_0$ are different for two cases [see Eqs.(\ref{281}) and (\ref{271})].

When energy  approaches the threshold $m$, the parameter $a\rightarrow \infty$.
Using Eq.(\ref{261}), Eq.(\ref{391}) and Eq.(\ref{401}) can be represented  by
\begin{align}\label{45}
&Im\{H^{(1)}_{1}(2\alpha_c)\}=0, \ for\ odd \ parity \ states,\notag\\
&Im\{H^{(1)}_{0}(2\alpha_c)\}=0,\ for\ even \ parity \ states,
\end{align}
where $Im\{H^{(1)}\}$ is the imaginary part of  $H^{(1)}$.
It should be emphasized that Eq.(\ref{45}) is exact result for $E=m$.

Further using the asymptotic formula of Hankel functions, i.e, $H^{(1)}_{\nu}(x)\sim \sqrt{\frac{2}{\pi x}}exp[i(x-\nu\pi/2-\pi/4)]$ as $x\rightarrow\infty$ , the critical potential strength $\alpha_c$ can be approximated by
\begin{align}\label{27}
&\alpha_c\simeq\frac{(-1/4+n)\pi}{2},\ for\ odd \ parity \ states,\notag\\
&\alpha_c\simeq\frac{(-3/4+n)\pi}{2},\ for\ even \ parity \ states,
\end{align}
where natural number $n=1,2,3,...$.

\textbf{Finally, it should be remarked that the above two different choices of boundary conditions at $x=x_0$, i.e. Eqs.(\ref{281}) and (\ref{271}),  correspond to two different self-adjoint extensions of Hamiltonian operator \cite{Bonneau2001}.
Further more, an arbitrarily  linear combination of above two kinds of  wave functions would form a new self-adjoint extensions of Hamiltonian operator. It is expected that the system would have a new energy spectrum which is an  interpolation of Fig.4 and Fig.5.  In this sense, the Hamiltonians with different boundary conditions would be different physical systems.}


\section{summary}
In conclusion, we investigate the bound states for a one-dimensional spin-1 Dirac model with a Coulomb-like potential of type III.
 We get the bound state energies by solving an effective Schr\"{o}dinger equation.
 It is found that the bound state energies can be well described by the quasi-classical approximation method.
Depending on the sign of potential strength over energy, i.e., $\alpha/E$, there exists two kinds of effective potentials.
For the case of  $\alpha/E<0$, there exists an infinite number of bound states near the threshold of upper continuous spectrum.
For large quantum number, the bound energy is consistent with the ordinary hydrogen atom bound state energy.

 For $\alpha/E>0$, similarly as that of Coulomb-like potential of type II, there also exists an infinite number of bound states which are generated from the flat band.
  When the bound state energies are very near the flat band, they are  proportional to the Coulomb-like potential strength $\alpha$ and form a $1/n$ energy spectrum.
 \textbf{We should emphasize that the existences of infinite bound states induced by potential are not limited to long ranged Coulomb potential.
Even for short-ranged potential, e.g., square well potential, there are also  infinite bound states.
The infinite bound states may also exist in the Cornel-like potential and the quark-antiquark interactions which usually have infinite values at infinities as required by quark confinements.}

 \textbf{ The above results would provide some useful insights in the understanding of flat band properties in many-body physics. For example, the infinite bound states induced by weak potential imply that an arbitrarily small interaction would dominate the physics.
 Since the bound states can appear for both repulsive and attractive potentials, then one can expect that even a repulsive interaction may result in superfluid/superconductor pairing states in flat band \cite{Kobayashi}.
   }    It is   expected that  $1/n$ energy spectrum may be observed experimentally in near future \cite{Wang2013,Mao2016}.
In addition, the existence of bound states in the continuous spectrum (BIC) needs further investigations.
%

\section*{Acknowledgements}
This work was supported by the NSFC under Grants Nos.
11874127.
%



\begin{thebibliography}{0}%
\makeatletter
\providecommand \@ifxundefined [1]{%
 \@ifx{#1\undefined}
}%
\providecommand \@ifnum [1]{%
 \ifnum #1\expandafter \@firstoftwo
 \else \expandafter \@secondoftwo
 \fi
}%
\providecommand \@ifx [1]{%
 \ifx #1\expandafter \@firstoftwo
 \else \expandafter \@secondoftwo
 \fi
}%
\providecommand \natexlab [1]{#1}%
\providecommand \enquote  [1]{``#1''}%
\providecommand \bibnamefont  [1]{#1}%
\providecommand \bibfnamefont [1]{#1}%
\providecommand \citenamefont [1]{#1}%
\providecommand \href@noop [0]{\@secondoftwo}%
\providecommand \href [0]{\begingroup \@sanitize@url \@href}%
\providecommand \@href[1]{\@@startlink{#1}\@@href}%
\providecommand \@@href[1]{\endgroup#1\@@endlink}%
\providecommand \@sanitize@url [0]{\catcode `\\12\catcode `\$12\catcode
  `\&12\catcode `\#12\catcode `\^12\catcode `\_12\catcode `\%12\relax}%
\providecommand \@@startlink[1]{}%
\providecommand \@@endlink[0]{}%
\providecommand \url  [0]{\begingroup\@sanitize@url \@url }%
\providecommand \@url [1]{\endgroup\@href {#1}{\urlprefix }}%
\providecommand \urlprefix  [0]{URL }%
\providecommand \Eprint [0]{\href }%
\providecommand \doibase [0]{http://dx.doi.org/}%
\providecommand \selectlanguage [0]{\@gobble}%
\providecommand \bibinfo  [0]{\@secondoftwo}%
\providecommand \bibfield  [0]{\@secondoftwo}%
\providecommand \translation [1]{[#1]}%
\providecommand \BibitemOpen [0]{}%
\providecommand \bibitemStop [0]{}%
\providecommand \bibitemNoStop [0]{.\EOS\space}%
\providecommand \EOS [0]{\spacefactor3000\relax}%
\providecommand \BibitemShut  [1]{\csname bibitem#1\endcsname}%
\let\auto@bib@innerbib\@empty
\end{thebibliography}%


\begin{thebibliography}{10}

\bibitem{Sutherland1986} Bill Sutherland, Localization of electronic wave functions due to local topology, Phys. Rev. B \textbf{34}, 5208 (1986).
\bibitem{Vidal1998} Julien Vidal, R\'{e}my Mosseri, and Benoit Doucot, Aharonov-Bohm Cages in Two-Dimensional Structures, Phys. Rev. Lett. \textbf{81}, 5888(1998).
\bibitem{Mukherjee} Sebabrata Mukherjee, et al., Observation of a Localized Flat-Band State in a Photonic Lieb Lattice, Phys. Rev. Lett. \textbf{114}, 245504(2015).

\bibitem{Mielke1999} Andreas Mielke, Ferromagnetism in Single-Band Hubbard Models with a Partially Flat Band, Phys. Rev. Lett. \textbf{82}, 4312(1999).
\bibitem{Zhang2010} Shizhong Zhang, Hsiang-hsuan Hung, and Congjun Wu, Proposed realization of itinerant ferromagnetism in optical lattices, Phys. Rev.  A \textbf{82}, 053618 (2010).
\bibitem{Raoux2014} A. Raoux, M. Morigi, J.-N. Fuchs, F. Pi\'{e}chon, and G. Montambaux,
From Dia to Paramagnetic Orbital Susceptibility of Massless Fermions,
Phys. Rev. Lett. \textbf{112}, 026402 (2014).
\bibitem{Leykam2017} Daniel Leykam, Joshua D. Bodyfelt, Anton S. Desyatnikov and Sergej Flach, Localization of weakly disordered flat band states. Eur. Phys. J. B \textbf{90}, \textbf{1} (2017).

\bibitem{Shen2010} R. Shen, L. B. Shao, Baigeng Wang, and D. Y. Xing, Single Dirac cone with a flat band touching on line-centered-square optical lattices, Phys. Rev. B \textbf{81}, 041410 (2010).

\bibitem{Urban2011} Daniel F. Urban, Dario Bercioux, Michael Wimmer, Wolfgang H\"{a}usler, Barrier transmission of Dirac-like pseudospin-one particles, Phys. Rev. B \textbf{84}, 115136 (2011).


\bibitem{Fang2016} A. Fang, Z. Q. Zhang, Steven G. Louie, and C. T. Chan, Klein tunneling and supercollimation of pseudospin-1 electromagnetic waves, Phys. Rev. B \textbf{93}, 035422 (2016).


\bibitem{Ocampo2017} Y. Betancur-Ocampo, G. Cordourier-Maruri, V. Gupta, and R. de Coss, Super-Klein tunneling of massive pseudospin-one particles, Phys. Rev. B \textbf{96}, 024304 (2017).
\bibitem{Yang2012} Shuo Yang, Zheng-Cheng Gu, Kai Sun, and S. Das Sarma, Topological flat band models with arbitrary Chern numbers,
Phys. Rev. B \textbf{86}, 241112(R) (2012).
\bibitem{Ghosh} Tutul Biswas and Tarun Kanti Ghosh, Dynamics of a quasiparticle in the $\alpha-T_3$
model: role of pseudospin polarization and
transverse magnetic field on zitterbewegung, J. Phys.: Condens. Matter \textbf{30}, 075301 (2018).

 \bibitem{Tovmasyan2018} Murad Tovmasyan, Sebastiano Peotta, Long Liang, P\"{a}ivi T\"{o}rm\"{a}, and Sebastian D. Huber, Preformed pairs in flat Bloch bands
Phys. Rev. B \textbf{98}, 134513(2018).

\bibitem{Volovik2019} Volovik, G.E. Flat Band and Planckian Metal. Jetp Lett. \textbf{110}, 352-353 (2019).

\bibitem{Peotta2015} S. Peotta,  and  P. T\"{o}rm\"{a},  Superfluidity in topologically nontrivial flat bands. Nat.Commun. \textbf{6}, 8944 (2015).
\bibitem{Cao2018} Yuan Cao, et.al., Unconventional superconductivity in magic-angle graphene superlattices, Nature \textbf{556}, 43 (2018).

\bibitem{Hazra2019} Tamaghna Hazra, Nishchhal Verma,and Mohit Randeria, Bounds on the Superconducting Transition Temperature: Applications to Twisted Bilayer Graphene and Cold Atoms, Phys. Rev. X \textbf{9}, 031049 (2019).



 \bibitem{Wuyurong2021} Yu-Rong Wu and Yi-Cai Zhang, Superfluid states in $\alpha-T_3$ lattice, Chinese Phys. B \textbf{30}, 060306 (2021).



















\bibitem{Julku2020} A. Julku, T. J. Peltonen, L. Liang, T. T. Heikkil\"{a}, and P. T\"{o}rm\"{a}
, Superfluid weight and Berezinskii-Kosterlitz-Thouless transition temperature of twisted bilayer graphene, Phys. Rev. B  \textbf{101}, 060505(R) (2020).
\bibitem{Kopnin2011} N. B. Kopnin, T. T. Heikkila, and G. E. Volovik
, High-temperature surface superconductivity in topological flat-band systems, Phys. Rev. B \textbf{83}, 220503(R) (2011).



\bibitem{Iglovikov2014} V. I. Iglovikov, et.al., Superconducting transitions in flat-band systems, Phys. Rev. B \textbf{90}, 094506 (2014).

\bibitem{Julku2016} Aleksi Julku, et. al.,  Geometric Origin of Superfluidity in the Lieb-Lattice Flat Band, Phys. Rev. Lett. \textbf{117}, 045303 (2016).

 \bibitem{Liang2017} Long Liang, et.al.,  Band geometry, Berry curvature, and superfluid weight, Phys. Rev. B \textbf{95}, 024515 (2017).

  \bibitem{Iskin2019} Iskin, M. Origin of fat-band superfuidity on the Mielke checkerboard lattice. Phys. Rev. A \textbf{99}, 053608 (2019).

\bibitem{Wu2021} Yu-Rong Wu, Xiao-Fei Zhang, Chao-Fei Liu, Wu-Ming Liu and Yi-Cai Zhang, Superfluid density and collective modes of fermion superfluid in dice lattice, Sci Rep \textbf{11}, 13572 (2021)


\bibitem{Economou} E. N. Economou, Green's Functions in Quantum Physics,(Springer-Verlag Berlin Heidelberg, Third Edition, 2006).

\bibitem{Zhangyicai2021} Yi-Cai Zhang, and Guo-Bao Zhu, Infinite bound states and hydrogen atom-like energy spectrum induced by a flat band, \url{https://www.researchgate.net/publication/354669549} (2021).

\bibitem{Gorbar2019} E. V. Gorbar, V. P. Gusynin, and D. O. Oriekhov, Electron states for gapped pseudospin-1 fermions in the field of a charged impurity, Phys. Rev. B \textbf{99} 155124(2019).

\bibitem{Pottelberge2020} R. Van Pottelberge, Comment on ``Electron states for gapped pseudospin-1 fermions in the field of a charged impurity", Phys. Rev. B \textbf{101}, 197102 (2020).
\bibitem{Han2019} Chen-Di Han, Hong-Ya Xu, Danhong Huang, and Ying-Cheng Lai, Atomic collapse in pseudospin-1 systems,  Phys. Rev. B \textbf{99}, 245413 (2019).
\bibitem{Zhangyicai20212} Yi-Cai Zhang, Wave function collapses and 1/n energy spectrum induced by a Coulomb potential in a one-dimensional flat band system, \url{https://www.researchgate.net/publication/355051219} (2021).






\bibitem{Zhang2013} Yi-Cai Zhang, Shu-Wei Song, Chao-Fei Liu, and Wu-Ming Liu, Zitterbewegung effect in spin-orbit-coupled spin-1 ultracold atoms, Phys. Rev. A \textbf{87}, 023612 (2013).

\bibitem{Huang2011} X Huang, Y Lai, ZH Hang, H Zheng, CT Chan, Dirac cones induced by accidental degeneracy in photonic crystals and zero-refractive-index materials, Nature Mater \textbf{ 10}, 582-586 (2011)

\bibitem{Chan2012} C. T. Chan, Zhi Hong Hang, and Xueqin Huang, Dirac Dispersion in Two-Dimensional Photonic Crystals, Adv. Optoelectron. 2\textbf{012}, 313984 (2012)









%
%
%
%








\bibitem{Downing2014} C. A. Downing and M. E. Portnoi, One-dimensional Coulomb problem in Dirac materials, Phys. Rev. A \textbf{90}, 052116 (2014).
\bibitem{Abramowitz}  Abramowitz, M. and Stegun, I. A. (Eds.). "Confluent Hypergeometric Functions." Ch. 13 in Handbook of Mathematical Functions with Formulas, Graphs, and Mathematical Tables, 9th printing. New York: Dover, pp. 503-515, 1972.
\bibitem{Wang1989} Z. X. Wang and D. R. Guo, Special Functions, (World
Scientific, Singapore, 1989).

\bibitem{Landau} L D Landau, E M Lifshitz, Quantum mechanics: non-relativistic theory, (Pergamon Press, Third Revised Edition, 1977).


\bibitem{Wang2013} Y. Wang, et al., Observing Atomic Collapse
Resonances in Artificial Nuclei on
Graphene, Science \textbf{340}, 734 (2013).

\bibitem{Mao2016} J. Mao, et al., Realization of a tunable artificial atom at a
supercritically charged vacancy in graphene, Nat. Phys. \textbf{12},545 (2016).








\bibitem{BIC} Yi-Cai Zhang, Bound states in the continnum (BIC) protected by self-sustained potentential barriers in flat band system,  \url{https://www.researchgate.net/publication/356223517}, (2021).
%
\bibitem{Bonneau2001} Guy Bonneau, Jacques Faraut, Galliano Valent, Self-adjoint extensions of operators and the teaching of quantum mechanics, American Journal of Physics \textbf{69}, 322 (2001). 


\bibitem{Kobayashi}
Keita Kobayashi, Masahiko Okumura, Susumu Yamada, Masahiko Machida, and Hideo Aoki, Superconductivity in repulsively interacting fermions on a diamond chain: Flat-band-induced pairing
Phys. Rev. B \textbf{94}, 214501(2016).





%
%
%
%
%
%


























%

%
%
%
%
%
%





%
%
%
%
%
%
%
%
%
%
%
%
%
%
%
%
%
%
%
%
%
%

%
%
%
%

%
%
%
%
%







%



















%









%
%
%
%
%
%
%
%
%
%









%
%
%



%
%
%



































\end{thebibliography}
\end{document}